\documentclass[letterpaper, 10 pt, conference]{ieeeconf}  % Comment this line out
\IEEEoverridecommandlockouts                              % This command is only
\usepackage{amsmath,amsfonts}
\usepackage[caption=false,font=normalsize,labelfont=sf,textfont=sf]{subfig}
\usepackage{stfloats}
\usepackage{url}
\usepackage{verbatim}
\usepackage{multirow}
\usepackage{cite}
\usepackage{xcolor}
\usepackage{mathtools}
\usepackage[capitalise]{cleveref}
\usepackage{cases}
\title{\LARGE \bf
Physics–Informed Neural Networks for Phase Locked Loop \\Transient Stability Assessment
}

\author{Rahul Nellikkath, Andreas Venzke, Mohammad Kazem Bakhshizadeh, Ilgiz Murzakhanov  and \\Spyros Chatzivasileiadis% <-this % stops a space
\thanks{This work is supported by the SYNERGIES project, funded by the European Commission Horizon Europe program, Grant Agreement No. 101069839, and by the ERC Starting Grant VeriPhIED, Grant Agreement No. 949899.}% <-this % stops a space
\thanks{R. Nellikkath, I. Murzakhanov and S. Chatzivasileiadis are with Department of Wind and Energy Systems, Technical University of Denmark (DTU), Kgs. Lyngby, Denmark 
        {\tt\small \{rnelli, ilgmu, spchatz\}@dtu.dk}}%
\thanks{A. Venzke and M. K. Bakhshizadeh are with Ørsted Wind Power, Nesa Alle 1, 2820, Denmark 
        {\tt\small \{anhve, modow\}@orsted.com}}%
}

\begin{document}

\maketitle
\thispagestyle{empty}
\pagestyle{empty}

%%%%%%%%%%%%%%%%%%%%%%%%%%%%%%%%%%%%%%%%%%%%%%%%%%%%%%%%%%%%%%%%%%%%%%%%%%%%%%%%
\begin{abstract}
A significant increase in renewable energy production is necessary to achieve the UN's net-zero emission targets for 2050. Using power-electronic controllers, such as Phase Locked Loops (PLLs), to keep grid-tied renewable resources in synchronism with the grid can cause fast transient behavior during grid faults leading to instability. However, assessing all the probable scenarios is impractical, so determining the stability boundary or region of attraction (ROA) is necessary. However, using EMT simulations or Reduced-order models (ROMs) to accurately determine the ROA is computationally expensive. Alternatively, Machine Learning (ML) models have been proposed as an efficient method to predict stability. However, traditional ML algorithms require large amounts of labeled data for training, which is computationally expensive. This paper proposes a Physics-Informed Neural Network (PINN) architecture that accurately predicts the nonlinear transient dynamics of a PLL controller under fault with less labeled training data. The proposed PINN algorithm can be incorporated into conventional simulations, accelerating EMT simulations or ROMs by over 100 times. The PINN algorithm's performance is compared against a ROM and an EMT simulation in PSCAD for the CIGRE benchmark model C4.49, demonstrating its ability to accurately approximate trajectories and ROAs of a PLL controller under varying grid impedance.
\end{abstract}

%%%%%%%%%%%%%%%%%%%%%%%%%%%%%%%%%%%%%%%%%%%%%%%%%%%%%%%%%%%%%%%%%%%%%%%%%%%%%%%%
\section{INTRODUCTION}
To achieve the net zero emission targets set by the UN for 2050, renewable energy production in the electricity grid has to be ramped up drastically in the coming decades. Unlike traditional synchronous power generators, the renewable resources such as Wind Power Plants (WPP) requires power-electronic controllers to remain in synchronism with the grid. In that respect, Phase Locked Loops (PLLs) are one of the most widely used controllers in renewable generators for tracking the grid reference frame \cite{ma2017}. However, recent studies have shown that in a weak grid, during large disturbances such as grid faults,  the PLL reference frame may fail to synchronize with the grid reference frame, leading to grid instability \cite{zhang}. Additionally, even during small perturbations, the interaction between the controller and a weak grid can result in small-signal instability, which is equally detrimental for the secure power supply. 

These types of complex and fast transient behaviors displayed by the grid-tied power-electronic controllers are challenging to examine using a traditional RMS analysis. Instead, power system operators must use EMT simulations to evaluate each scenario. However, EMT simulations are computationally very expensive, which makes it is impractical to assess each scenario on a case-by-case basis. Instead, power system operators usually rely on a pre-determined stability boundary of the controller, also called Region of Attraction (ROA), which guarantees a domain of safe operation. More specifically, ROA is a space of initial system states within which all trajectories will converge to a stable equilibrium point. 
%Thus with the additional nonlinearity in the network, it is crucial to map the renewable generator's stability boundary or determine the ROA before it can be deployed on an existing network. 

However, to accurately determine the ROA using EMT simulation for a renewable generation plant, the power system operators still have to evaluate the stability of numerous scenarios to cover all the parameter combinations. Evaluating all the necessary scenarios to determine the ROA using EMT simulations is computationally intractable. Therefore, it is essential to develop methods to assess the stability of these controllers.

This has led to the development of reduced-order models (ROMs) and approximations of the actual converter dynamics. The most simple are the linearised model-based approaches, such as eigenvalue analysis \cite{ma,hu} or impedance-based stability analysis \cite{liu,amin}, that approximates the wind turbines (WT) grid side converter and the connected power system with a linear function around an operating point. These models perform well at analyzing the stability under small disturbances but cannot sufficiently capture the instability under larger grid disturbances. Hence, a nonlinear approach must be used to predict global asymptotic stability  \cite{vu,xu,Kazem_ROM}. 

When it comes to non-linear reduced order modelling approaches, even though the once proposed are considerably faster than the EMT simulations, they still demand a long computational time to evaluate the stability at all the points of interest. An alternative option explored in \cite{Lyp} is to use Lyapunov's direct method to estimate the ROA. However, as shown in \cite{sujay2}, Lyapunov's direct method is a conservative approximation of the ROA and could falsely classify multiple stable initial states as unstable. 

Hence, developing a more efficient method than EMT simulations and reduced-order models is essential to accurately assess the stability of renewable generation plants. To this end, in \cite{ml1}, Machine Learning (ML) models have been proposed to predict the stability of WT. ML models can deliver solutions 100-1'000 faster than conventional models with sufficient accuracy  \cite{transient}. Thus they can be used to quickly screen a vast number of scenarios to approximate the Region of Attraction, and identify few critical scenarios that require further analysis with EMT simulations. 
%quickly to identify critical scenarios that are further analysed with the full EMT simulations. 
 
 Still traditional ML algorithms, such as Neural Networks (NN), require large amounts of high quality labeled data for training. Generating such a training dataset requires substantial computation time, which would cancel out the speedup the ML algorithms could offer. The development of physics-informed NNs (PINNs) addresses exactly this challenge, as it incorporates the underlying physics inside the neural network training, and by that, it drastically reduces the dependency of NN performance on external training data. As we will see later in this paper, this fundamental property of PINNs eliminates the need to generate large training datasets, avoiding a heavy computational burden, and drastically accelerates PINN training \cite{PINN1}. 

PINN for predicting the transient stability of an equivalent 2-area transmission grid with synchronous machines was proposed in \cite{transient}. Our previous work has also looked into using PINNs to evaluate a wind farm's N-1 small-signal stability margin \cite{closing}. 
%However, these methods have yet to accurately approximate the nonlinear transient dynamics of a power electronic controller during fault. 
This is the first paper that uses Physics-Informed Neural Networks (PINNs) to approximate the dynamics of Electro-Magnetic Transient Simulations. We believe that deploying PINNs for so computationally expensive simulations as EMT puts forward the most valuable use of PINNs, where we can achieve a speedup of over 100 times compared to conventional methods (including the time taken for training the PINNs), while maintaining sufficient accuracy. As a use case, in this paper, we focus on the accurate approximation of the non-linear power system dynamics of a PLL controller under fault.   

\begin{enumerate}
    \item We demonstrates that the proposed PINN architecture can accurately estimate the Region of Attraction of a PLL controller with varying grid strength more than 100 times faster than a Reduced-Order Model.
    \item We introduce a novel recurring PINN architecture that can provide accurate predictions regardless of the prediction time window. In other words, PINNs can accurately learn the underlying physics of a system using only a narrow prediction window of e.g. 100-200 ms. We can then use this learnt neural network model in a recurrent fashion to predict for much longer horizons, e.g. 2s. This is an important contribution since it allows for more flexible and robust predictions in real-world scenarios.
\end{enumerate}
\begin{comment}
    This paper proposes a PINN training architecture to accurately predict the non-linear system dynamics of a PLL controller under fault. Our contributions are as follows:
\begin{enumerate}
\item We use the second-order ROM for PLL stability proposed in \cite{Kazem_ROM} to develop a PINN capable of estimating the non-linear transient system dynamics of a PLL controller during fault.
\item We introduce a novel recurring PINN architecture that can provide accurate predictions regardless of the prediction time window. This is an important contribution since it allows for more flexible and robust predictions in real-world scenarios where the time taken for the controller to reach stable equilibrium may vary.
\item Our experiment demonstrates that the proposed PINN architecture can accurately estimate the stability boundary or ROA of a PLL controller with varying grid strength in a fraction of the time it would take the ROM to do so.
\end{enumerate}
\end{comment}

The remainder of this paper is structured as follows: Section II describes the ROM for a PLL controller global stability analysis. Section III introduces the PINN algorithm used to approximate the non-linear system dynamics. Section IV presents the simulation setup and the results demonstrating the performance of the proposed PINN training architecture. Section IV concludes. 
\section{Reduced Order Model for PLL control}
To develop a PINN capable of estimating the non-linear transient system dynamics of a PLL controller during fault, we use the second-order Reduced Order Model (ROM) for PLL stability proposed in \cite{Kazem_ROM} and shown in Fig. 1. The ROM in \cite{Kazem_ROM} was developed for a renewable generator such as WT with PLL controller under fault and is actively used in industry during wind farm design. Once the fault occurs, the dc chopper is activated, which allows to assume the dc voltage that the grid-side converter (GSC) receives as constant. Considering, besides that, the inner current controller is assumed to be fast enough to regulate the dq-domain currents during fault, the dynamics of the inner current loop can be neglected. This assumption is valid since fast system dynamics could be neglected while analyzing slow PLL dynamics of the GSC. The electrical model of the renewable generator, here a WT, after the dc chopper is activated is given in Fig. 1. 
\begin{figure}
  \centerline{\includegraphics[scale=.35]{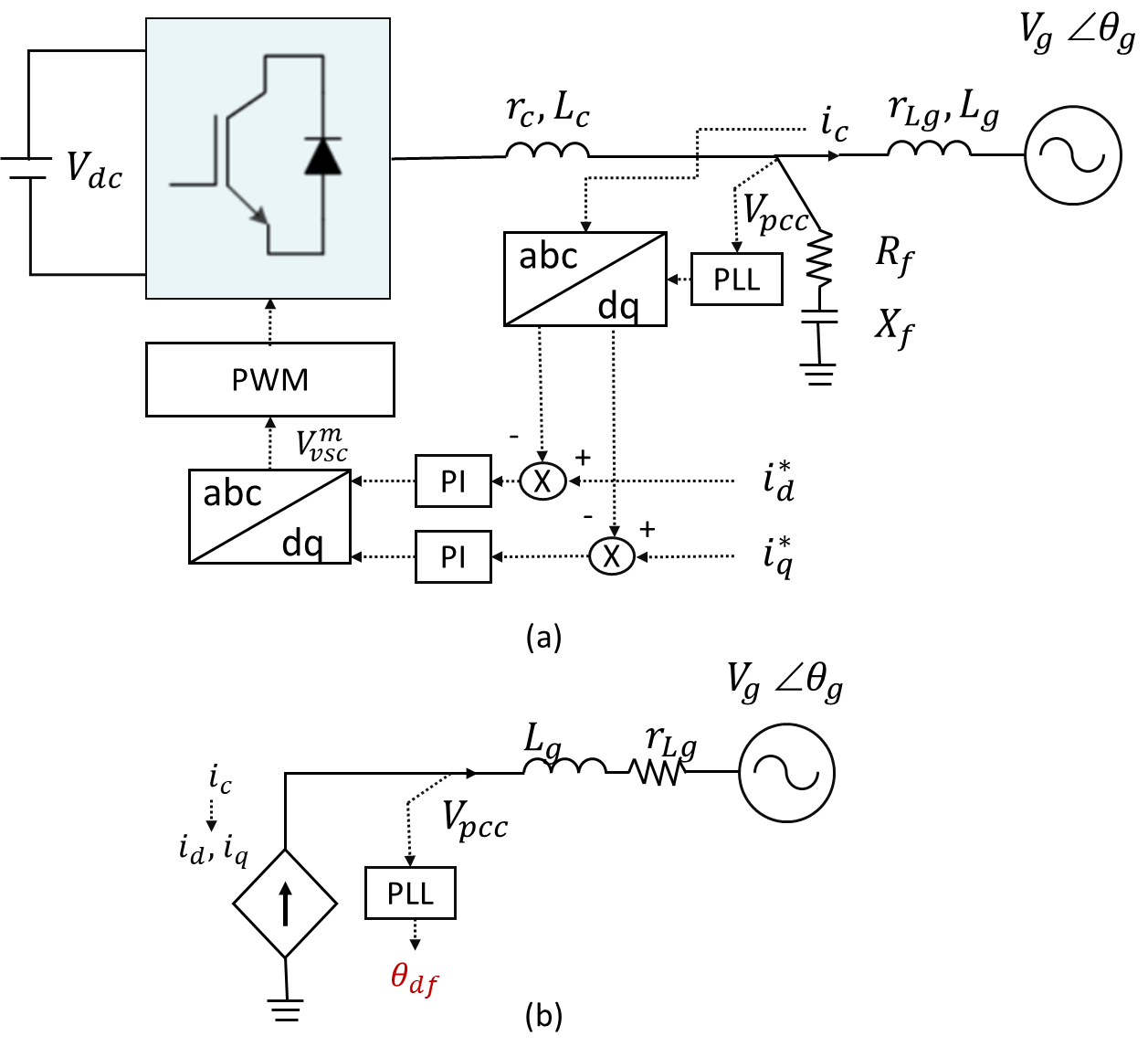}}
  \caption{(a) The electrical model of the renewable generator with PLL controller with grid-side converter and controls. (b) The reduced-order electrical model of renewable generator.}
\end{figure}

As shown in Fig. 1b, while developing the ROM, the shunt capacitor of the filter, denoted by $X_f$ in  Fig. 1a, was neglected since it had negligible impact on grid synchronisation stability when the current is controlled on the grid-side inductor of the LCL filter \cite{Kazem_ROM}. The resulting reduced-order representation of a renewable generator and grid in the dq domain is illustrated in Fig. 2a

\begin{figure}
  \centerline{\includegraphics[scale=.3]{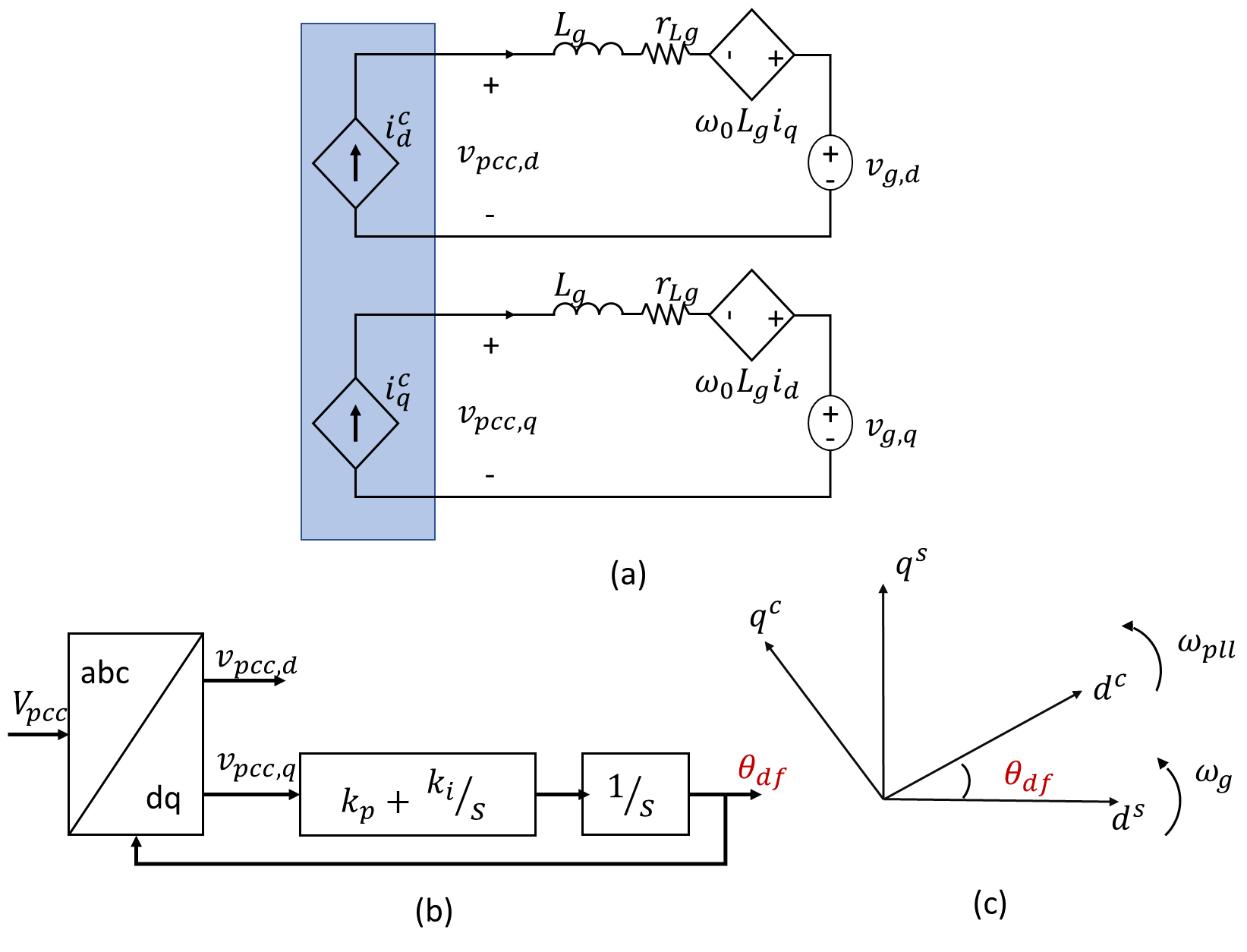}}
  \caption{(a) Reduced-order WT system representation in the dq domain (b) Typical synchronous reference frame PLL. (c) Vector diagram: Misalignment between the PLL reference frame and the grid reference frame.}
\end{figure}

The reduced order model was proposed for a PLL with a synchronous rotation frame (SRF) approach, which identifies the utility grid voltage phase angle by synchronizing the rotating frame reference of the PLL with the grid voltage. The SRF-PLL to track the grid voltage phase angle is given in Fig. 2b, where $\theta_{df}$ is the angle tracked by the PLL. The grid frame rotates with the grid frequency $\omega_g$, and the PLL reference frame rotates with the PLL frequency $\omega_{pll}$. The resulting misalignment between the grid voltage angle, denoted by $\theta_g$, and $\theta_{df}$, is depicted in Fig. 2c.

Assuming time-invariant system parameters after clearing the fault, the SRF-PLL transient nonlinear dynamics can be formulated as follows (see \cite{Kazem_ROM} and \cite{sujay} for the derivations):

\begin{gather}
   \theta_{df} = \int ( k_p \cdot v_{pcc,q}^c + k_i \int v_{pcc,q}^c dt) dt \\
   \begin{split}
   v_{pcc,q}^c = -V_g \cdot sin(\theta_{df} - \theta_g) + r_{Lg} i_q^c + L_g i_d^c \omega_g  \\  + L_g i_d^c (\dot{\overline{\theta_{df} - \theta_g}})
   \end{split} \label{ROM1}
\end{gather}

where $k_p$ and $k_i$ are the control parameters of the SRF-PLL. $v_{pcc,d}$ and $v_{pcc,q}$ are d and q coordinates of the voltage at the point of common coupling ($V_{pcc}$) in the dq reference frame. $V_g$ is the grid voltage, and $r_{Lg}$ and $L_g$ denotes the grid side impedance. $i^c_d$ and $i^c_q$ are the dq currents in the PLL reference frame. 

The second-order dynamics in \eqref{ROM1} can be transformed into an equivalent swing equation of the PLL controller as formulated in \cite{Kazem_ROM}:

\begin{equation}
\begin{bmatrix}
\dot{\delta} \\
M \dot{\omega}
\end{bmatrix} =
\begin{bmatrix}
\omega \\
T_m - T_e -D \omega 
\end{bmatrix} 
\end{equation}

\begin{align}
    \begin{split}
        M &= 1-k_p L_g i_d^c \\
        T_m &= k_i(r_{Lg} i_q^c + L_g i_d^c \omega_g) \\
        T_e &= k_iV_g sin(\delta) \\
        D &= k_p V_g cos(\delta) - k_iL_gi_d^c
    \end{split}
    \label{ROM_swgn}
\end{align}

where, $\delta = \theta_{df} - \theta_g$ is the misalignment in the angle tracked by the PLL. By considering a vector $x = [ \delta, \omega]^T$, we can represent the system dynamics in a more compact form as follows:

\begin{equation}
    \frac{d}{dt}x = f(t,x,u)
\end{equation}

where $x$ is the state of the system, and $u$ is the system and control parameters. For a given initial conditions $x_0$ and system parameter $u$, we can use an ODE solver to evaluate trajectory by performing an integration over small time steps. However, this can be computationally intensive, particularly when we need to integrate over a long time horizon and for multiple initial conditions and parameters. To address this challenge, we propose a physics-informed neural network (PINN) that can accurately approximate the trajectory $x$ with significantly less computational resources.

\section{Physics-Informed Neural Network to Approximate the Reduced Order Model}
Neural Networks (NNs) are considered global approximators. A NN of sufficient size, which has been appropriately trained is guaranteed to be able to determine the output of any function, including those of the ROM we consider in this paper, without loss of accuracy. To achieve this, the NN uses a group of interconnected hidden layers with multiple neurons to learn the relationship between the input and output layers. For our problem, the inputs to the NN are the state of the system after the fault is cleared ($x_0$), the prediction time, i.e. the time at which the NN should make the prediction (denoted by $t$), and a few of the relevant system parameters such as grid impedance (indicated by $u$). The NN to approximate the PLL state at time $t$ can be depicted as follows:

\begin{equation}
    x \approx \hat{x} = NN(x_0,t,u)
\end{equation}

\begin{figure}
\centerline{\includegraphics[scale=.4]{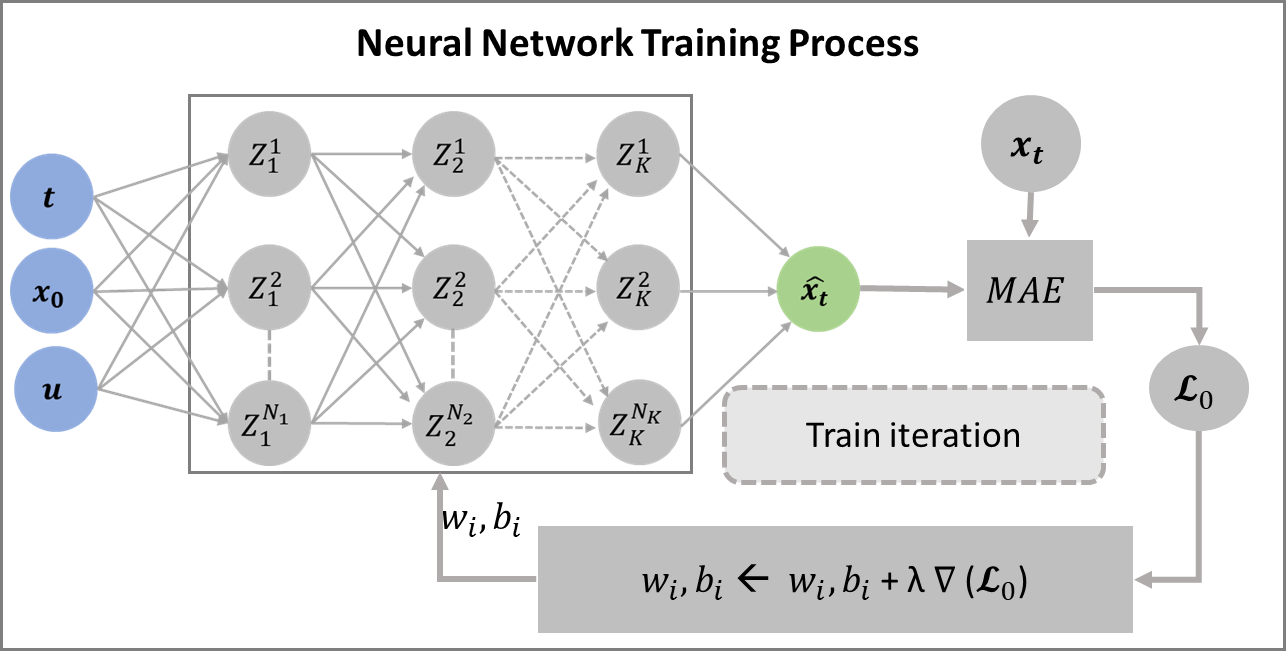}}
\caption{Illustration of the neural network architecture to predict the evolution of the reduced-order model: There are K hidden layers in the neural network with $N_k$ neurons each. Where k = 1, ...,K.}
\label{NN_basic}
\end{figure}

A standard NN, with $K$ number of hidden layers and $N_k$ number of neurons in hidden layer $k$, is shown in \cref{NN_basic}. Each neuron in a hidden layer is connected with neurons in the neighboring layers through a set of edges. The information exiting one neuron goes through a linear transformation over the respective edge before reaching the neuron in the subsequent layer. In every neuron, a nonlinear so-called "activation function" is applied to the information to introduce nonlinear relationships into the approximator. The NN for approximating the trajectory $x$ can be formulated as follows:

\begin{align}
    {Z}_0 &= [t,x_0,u] \\
    \hat{\mathbf{Z}}_k &= \mathbf{w_{k}}\mathbf{Z}_{k-1}+\mathbf{b_{k}}\label{NN1} \\
    \mathbf{Z}_k &= \sigma( \hat{\mathbf{Z}}_k ) \label{eq:sigma}\\
    \hat{x} &=  \mathbf{w_{K}}\mathbf{Z}_{K-1}+\mathbf{b_{K}}
\end{align}

where ${\mathbf{Z}}_{k}$ is the output of the neurons in layer $k$, $\hat{\mathbf{Z}}_k$ is the information received at layer $k$, $\mathbf{w_{k}}$ and $\mathbf{b_{k}}$ are the weights and biases connecting layer $k-1$ and $k$, and $\sigma$ is the nonlinear activation function. There is a range of possible activation functions, such as the sigmoid function, the hyperbolic tangent, the Rectifier Linear Unit (ReLU), and others. In this paper, we use the ReLU as the activation function, similar to the vast majority of recent papers, as this has been shown to accelerate the NN training \cite{glorot}. The ReLU activation function can be formulated as follows:

\begin{align}
    \mathbf{Z}_k &= \max( \hat{\mathbf{Z}}_k,0)\label{Relu}
\end{align}

The average error in predicting the state of the PLL (denoted by $\mathcal{L}_{0}$) for different starting points and time steps in the training data set is measured by:

\begin{equation}
    \mathcal{L}_{x} = \frac{1}{N} \sum_{i=1}^N | \mathbf{x}_{i} - \hat{\mathbf{x}}_{i}| \label{L_0}
\end{equation}

where $N$ is the total number of data points in the training set, $\mathbf{x}_i$ is the state of the system obtained by solving the ODE, and $\hat{\mathbf{x}}_i$ is the predicted state of the system.

The NN is trained using the backpropagation algorithm, which modifies the weights and biases of the NN in every iteration of the NN training to minimize the average prediction error $\mathcal{L}_{0}$. However, for the NN to accurately approximate the dynamics of the nonlinear system, it will require a considerable amount of training data. Creating this amount of training data is computationally expensive and can render this approach not feasible. To overcome this challenge, we propose a physics-informed neural network (PINN) to approximate the reduced-order model in this paper. 

\subsection{Physics-Informed Neural Network}
Considering the NN prediction should satisfy the ROM formulation given in \eqref{ROM_swgn}, to improve the NN generalization capabilities, we can impose that the temporal derivative of the NN's approximation $\hat x$ w.r.t the time $t$ (i.e., $\frac{d}{dt} \hat x$),  calculated using automatic differentiation (AD), matched the state update f (t, x,u) at all the training data points. This can be promoted by including the following loss function in the NN training:

\begin{equation}
    \mathcal{L}_{dt} = \frac{1}{N} \sum_{i=1}^N | \mathbf{f (t, x_i,u)} - {\mathbf{\frac{d}{dt} \hat x _{i}}}| \label{L_df}
\end{equation}

The resulting NN (denoted by dtNN) loss function can be formulated as follows:

\begin{equation}
   \mathcal{L}_{dtNN} =  \Lambda_x \mathcal{L}_x + \Lambda_{dt} \mathcal{L}_{df}
   \label{L_dt}
\end{equation}

where $\Lambda_x$ and $\Lambda_{dt}$ are the weights given to the respective loss functions. 

Furthermore, we can assess the accuracy of the NN's prediction $\hat x$ by comparing its temporal derivative $\frac{d}{dt}\hat x$ with the state update computed by the neural network approximation $f(t,\hat x,u)$. This way, we can consider additional training points, also called collocation points, in the training space for which we do not have to use computational resources to evaluate the ODE solution $x$. Instead, the reduced order model will help NN learn the dynamics of the controller for these new training data points. The loss function can be formulated as follows:

\begin{equation}
    \mathcal{L}_{f} = \frac{1}{N} \sum_{i=1}^N | \mathbf{f (t, \hat x_i,u)} - {\mathbf{\frac{d}{dt} \hat x _{i}}}| \label{L_f}
\end{equation}

\begin{figure}
\centerline{\includegraphics[scale=.37]{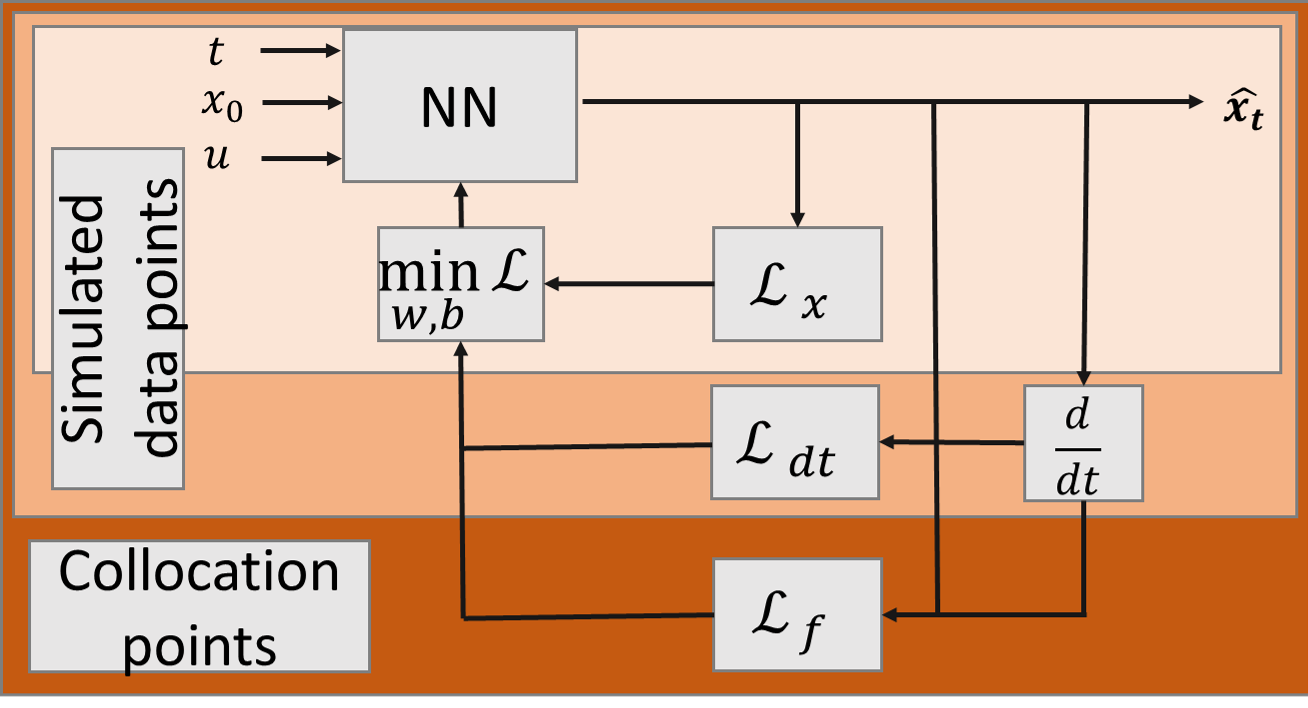}}
\caption{Illustration of the Physics-informed neural network architecture to predict the evolution of the reduced-order model: There are K hidden layers in the neural network with $N_k$ neurons each. Where k = 1, ...,K.}
\label{PINN_fig}
\end{figure}

The loss functions given in \eqref{L_df} and \eqref{L_f} can be weighted and added to the NN loss function in \eqref{L_dt} to get the proposed PINN loss function as follows:

\begin{equation}
   \mathcal{L} =  \Lambda_x \mathcal{L}_x + \Lambda_{dt} \mathcal{L}_{df} + \Lambda_{f} \mathcal{L}_{f}
\end{equation}

where $\Lambda_{f}$ is the weights given to the loss functions $\mathcal{L}_{f}$. The proposed PINN algorithm for training the NN is given in \cref{PINN_fig}.

\subsection{Recurrent PINN for PLL stability assessment}
NN for predicting time series data usually performs much better when trained for a short time window. However, the time required for the PLL to reach a stable equilibrium point depends on the initial values of $\delta$ and $\omega$ after the fault is cleared. The initial conditions closer to the equilibrium point could reach the equilibrium much faster than the ones much further. This implies that the NN approximator would have to work for a wide range of prediction time $t$ to make accurate predictions about the stability of the PLL. However, considering the underlying dynamics of the PLL controller (the ROM in \eqref{ROM_swgn}) remains the same until the system reaches equilibrium. Therefore, designing a huge NN to approximate the trajectory for a wide range of prediction time windows would be inefficient.

\begin{figure}[ht]
    \centerline{\includegraphics[scale=.35]{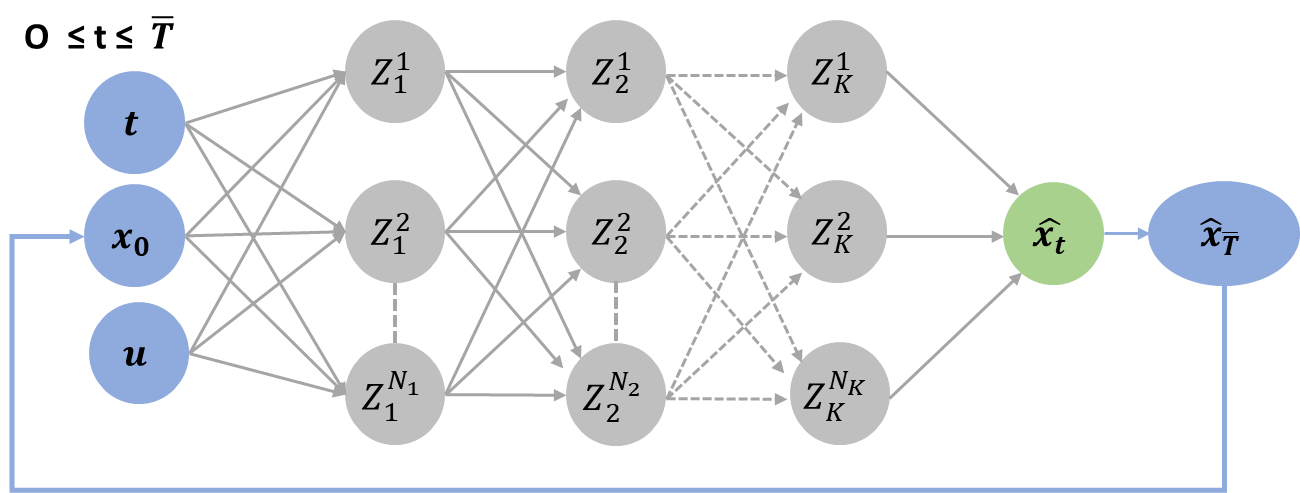}}
    \caption{Illustration of the recurrent physics-informed neural network (Re-PINN) architecture to predict the evolution of the reduced-order model: Re-PINN approximate the system state for a time t $\in$ [0, $\overline{T}$]. If the system did not reach equilibrium in that interval, then we can give the state of the system at $\overline{T}$, denoted by x($\overline{T}$), as starting point to the Re-PINN and evaluate again.}
    \label{REC_PINN}
\end{figure}

To address this issue, we developed a recurrent PINN, denoted by Re-PINN, by limiting the prediction window of the NN approximator to a fixed value, denoted by $\overline{T}$. Then, once deployed, we use the Re-PINN to approximate the system state for a time t $\in$ [0, $\overline{T}$]. If the system did not reach equilibrium in that interval, then we can give the state of the system at $\overline{T}$ (i.e. x($\overline{T}$)), as starting point to the Re-PINN and evaluate again. This allows the Re-PINN, as depicted in \cref{REC_PINN}, 
to make accurate predictions regardless of the time it takes for the PLL controller to reach stability.

Limiting the prediction window of the NN to a fixed value, $\overline{T}$,  helped us reduce the NN size while maintaining the same level of accuracy. Moreover, for a small value of $\overline{T}$, this approach also allowed us to restrict the training dataset time window to $\overline{T}$, thereby reducing the computational time required to solve the ROM for generating the training dataset. 

However, an extremely small value of $\overline{T}$ could make the nonlinear dynamics of the PLL controller appear linear to the NN approximation, resulting in a possible piece-wise linear approximation of the PLL controller dynamics. Furthermore, with an extremely small value of $\overline{T}$, the user will have to make multiple NN predictions to approximate the entire trajectory of PLL till it reaches the equilibrium. Hence, it is essential to choose a suitable value of $\overline{T}$ based on the system dynamics. In the case studies discussed in \cref{Resultsandcase}, we used a $\overline{T}$ of 100ms. 
\section{Results} \label{Resultsandcase}
This section presents a comparative analysis of the proposed Re-PINN approach with a recurring NN (denoted as Re-NN) and Re-dtNN, which compares the temporal derivative of the NN with the reduced-order model (ROM) introduced in \cite{Kazem_ROM} to accurately predict the dynamics of a PLL controller. The ROM is known to capture the slow dynamics of a PLL controller accurately \cite{Kazem_ROM}. Hence, we analytically compare the performance of the Re-PINN approximator, the Re-NN approximator, and the Re-dtNN approximator against the ROM. Furthermore, the trajectory predicted by Re-PINN is also compared to an EMT switching simulation model of a WT in PSCAD on the CIGRE benchmark model C4.49, using the system and control parameters provided in \cref{sandc}.

\begin{table}
\caption{SYSTEM AND CONTROL PARAMETERS}
\centering
\label{sandc}
\begin{tabular}{lll}
\hline \hline
Symbol        & Description                                                                                  & Value        \\ \hline \hline
$S_b$         & Rated Power                                                                                  & 12 MVA       \\
$V_g$         & Nominal grid voltage (L-N, pk)                                                               & 690 V        \\
$V_{dc}$      & DC-link voltage                                                                              & 1.38 kV      \\
$f_0$         & Rated frequency                                                                              & 50 Hz        \\
$T_s$         & Simulation time step                                                                         & 10 $\mu s$ \\
$r_{Lg},L_g$  & Grid-side inductor (pu)                                                                      & X/R = 16.3   \\
$i^c_d,t^c_q$ & \begin{tabular}[c]{@{}l@{}}Pre-disturbance active and \\ re-active current (pu)\end{tabular} & 1.0, -0.1    \\
$K_{cc}$      & Current contoller gains                                                                      & 0.05, 0.3    \\
$K_{pll}$     & SRF PLL gains                                                                                & 0.025, 1.5  \\ \hline \hline
\end{tabular}
\end{table}
\subsection{Neural Network Training Setup}
The objective of this work was to analyzes the ability of Re-PINN to accurately predict the transient stability of the system under varying grid strength. Thus, to simplify the training process, all system and control parameters other than the grid impedance were assumed constant while generating the dataset. Furthermore, the X/R ratio of the grid was considered to be constant. Then while generating the training and test dataset, the value of both $r_{Lg}$ and $L_g$ was changed by a factor of $\alpha \in [0.1, 2]$. The initial state space for $\delta$ and $\omega$ was limited to $-\pi$ to $\pi$ radians and -60 to 60 radians/second, respectively. We chose this region to limit the number of unstable initial states since an unstable initial state typically results in a high value of $\omega$ and a rapidly varying $\delta$. Accurately capturing the dynamics of such states would require a large NN. Moreover, the controller is usually cut off after a specific cut-off frequency $\omega$ for unstable initial states, making accurate predictions of their system state less critical. Thus, it would be more crucial if the NN could accurately predict the dynamics of stable system well and is able to identify unstable initial states.

To train the three ML algorithms, we used a generated training dataset that comprised 12,000 independent random trajectories spanning 100 ms, implemented using the Runge–Kutta solver in SciPy. In addition, we provided the proposed Re-PINN model with 24,000 random initial states from the input domain as collocation points to enhance the accuracy of the predictions. Unlike the training dataset, we did not compute the system trajectory for these collocation points. To evaluate the performance of the three ML algorithms, we tested them on a dataset of 24,000 independent random trajectories spanning 1 second

A Re-NN, Re-dtNN, and Re-PINN with four hidden layers and 100 nodes in each layer are used to predict the ROM solutions. The ML algorithms were implemented using PyTorch. WandB \cite{wandb} was used for monitoring and tuning the hyperparameters. The NNs have trained in a High-Performance Computing (HPC) server with an Intel Xeon E5-2650v4 processor and an NVIDIA Tesla V100 GPU with 16 GB RAM. The code and datasets to reproduce the results are available online \cite{Code_git}.

\subsection{Comparing Different Neural Network Algorithms}

This section evaluates the performance of the three NN models in predicting the dynamics of PLL under fault. The aim was to assess the generalization capabilities of each ML algorithm by comparing the NN predictions with the ROM solutions in the test set. The mean absolute error (MAE) in the test set was recorded during the NN training iterations, as shown in \cref{Res_NNvPINN}.
\begin{figure}
  \centerline{\includegraphics[scale=.6]{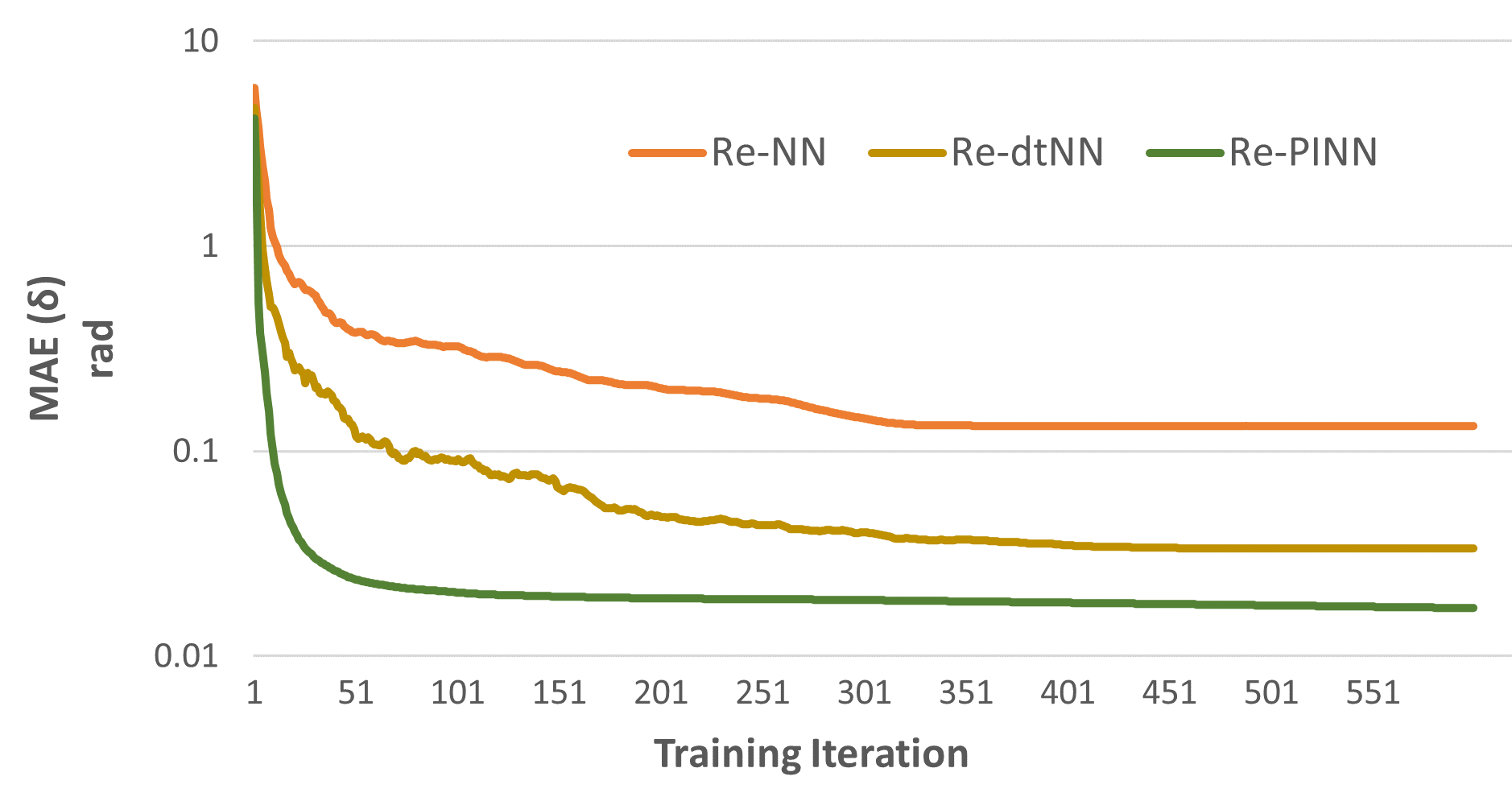}}
  \caption{The mean absolute error (MAE) in the test set during NN training iterations for different NN algorithms.}
  \label{Res_NNvPINN_MAE}
\end{figure}

The results indicate that the Re-dtNN model, which compared the temporal derivative of the NN with ROM in the training dataset, achieved significantly better prediction accuracy in the test set than a similarly sized Re-NN model. Moreover, the Re-PINN model outperformed the Re-NN and Re-dtNN models by almost an order of magnitude, thanks to the additional collocation points in the training set. Furthermore, with these extra collocation points, the Re-PINN model converged much faster than the other two NN models.

Additionally, to investigate if the Recurrent NN models resulted in error propagation as prediction time increases, we plotted the MAE in predicting the trajectory as a function of prediction time for all stable initial states in the test set. The MAE of Re-NN, Re-dtNN, and Re-PINN in predicting the PLL state $\delta$ was compared against the ROM at 50 fixed time steps between zero and one second, as illustrated in \cref{Res_NNvPINN}.
\begin{figure}
  \centerline{\includegraphics[scale=.55]{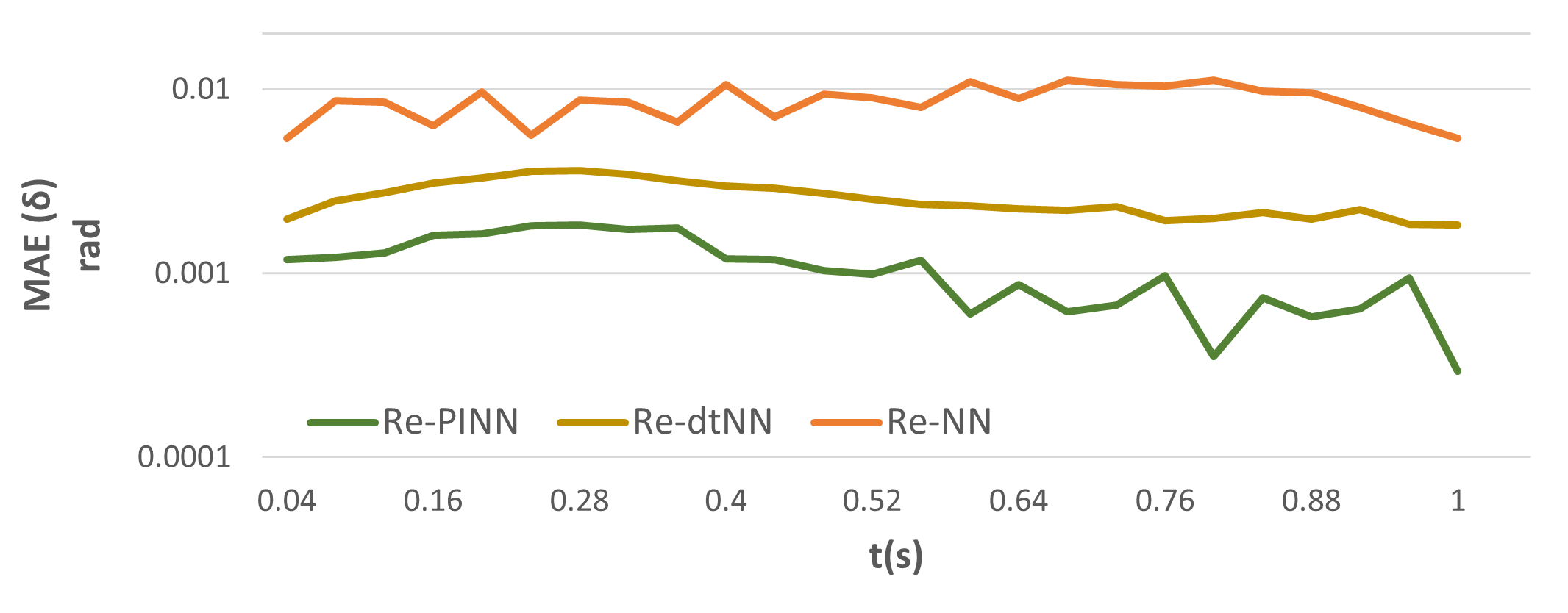}}
  \caption{Comparing the performance of the trained NN algorithms.}
\label{Res_NNvPINN}
\end{figure}

The results indicate that none of the three NN models suffer from error propagation in the test set. Additionally, it was observed that the Re-PINN predictions were almost an order of magnitude more accurate than the Re-NN model regardless of the prediction time, and with Re-PINN, the MAE reduced as the prediction time increased.

\begin{figure}
  \centerline{\includegraphics[scale=.25]{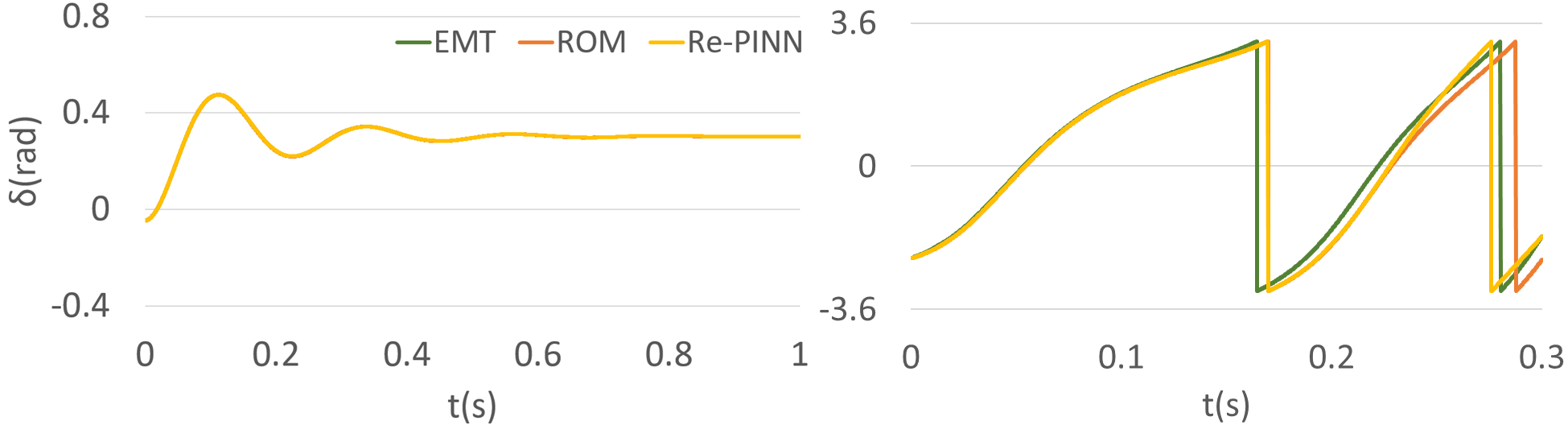}}
  \caption{Trajectory predicted by the Re-PINN as compared to ROM and PSCAD for an initial grid voltage phase jump of $20^0$ (left) and $150^0$ (right).}
\label{Res_Stable}
\end{figure}

Furthermore, the trajectory predicted by the Re-PINN was also compared against an EMT switching simulation in PSCAD and Re-PINN managed to capture the dynamics reasonably well for a stable and an unstable initial state as shown in \cref{Res_Stable}.  

Additionally, our experiments revealed that the dataset generation for training and testing required half an hour, while training all three NN architectures using GPU took only between 10 to 15 minutes each. These results indicate that dataset creation is a more time-consuming task than NN training. This shows that the PINN architecture achieved significantly better results compared to the other NN architectures while requiring a comparable amount of computational time.

\subsection{Predicting Region of Attraction using Physics-Informed Neural Network}
To ensure that Re-PINN did not misclassify any unstable cases as stable, we compared the region of attraction (ROA) predicted by Re-PINN against the ROA obtained analytically using ROM solutions. ROA is the set of initial states of the system from which all trajectories will converge to a stable equilibrium point. ROA can be calculated using Lyapunov's direct method or equal area criteria. However, these methods are often approximations and do not accurately capture the region \cite{sujay2}. In this section, we demonstrate the ability of PINNs to accurately predict the ROA for a PLL controller in CIGRE benchmark model C4.49 by evaluating the trajectory at multiple initial states. 

\begin{figure*}
  \centerline{\includegraphics[scale=.35]{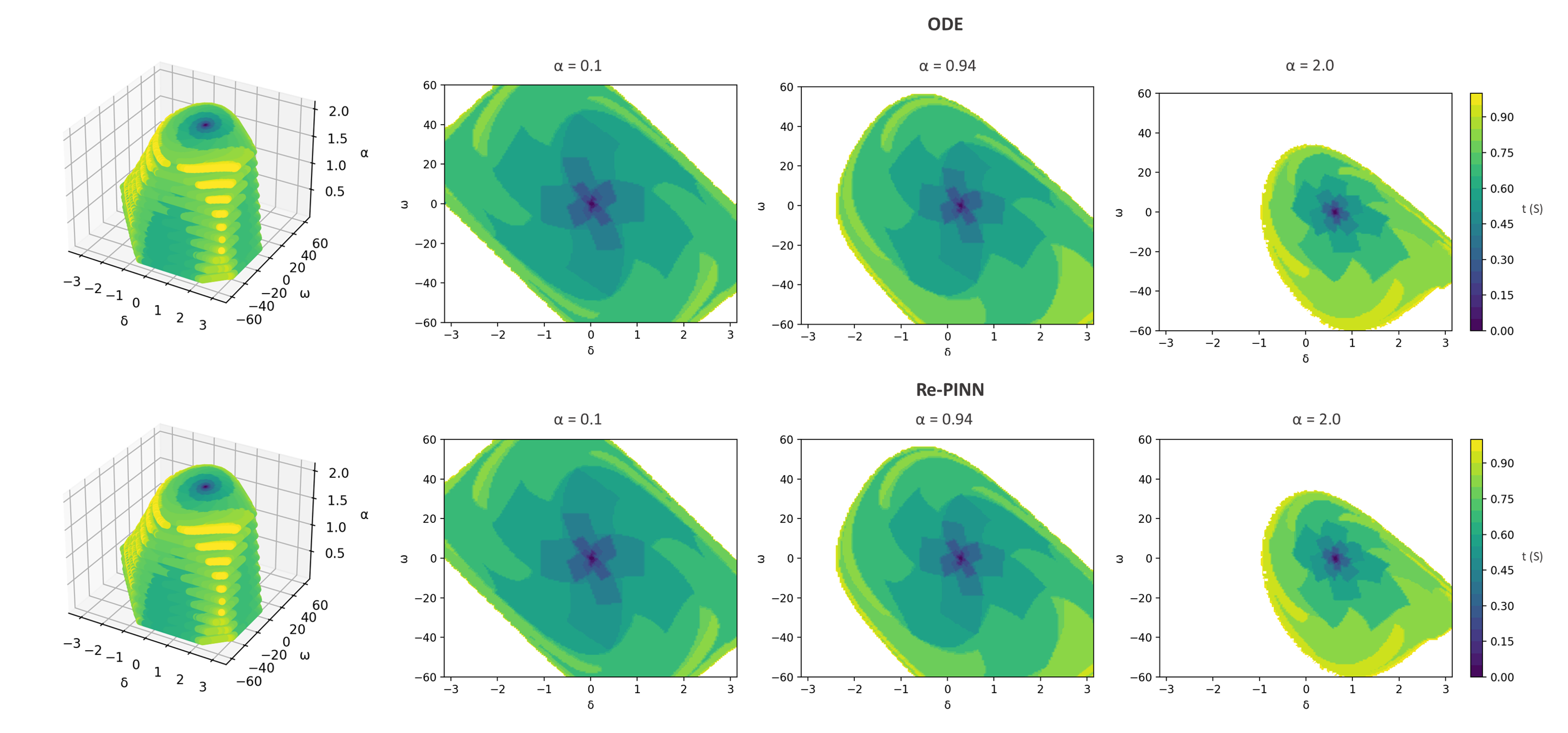}}
  \caption{Region of attraction predicted by the Re-PINN and the ROM}
  \label{Res_ROA}
\end{figure*}

The contour plot of the ROA with the time taken for the initial condition to reach a stable equilibrium is shown in \cref{Res_ROA}. We calculated the time taken to reach the stable equilibrium as follows:
\begin{equation}
     t= t_{eq}
\end{equation}
s.t.
\begin{align}
    \delta (t_{eq}) &= \delta_{eq} \pm \epsilon_\delta \\
    \omega (t_{eq}) &= \omega{eq} \pm \epsilon_\omega 
\end{align}

where $\delta_{eq}$ and $\omega{eq}$ are the equilibrium states that can be achieved after the fault is cleared, they are computed using the formulation given in \cite{Kazem_ROM}. $\epsilon_\delta$ and $\epsilon_\omega$ are small error terms used to negate the tiny fluctuations in the Re-PINN prediction. The ROA was evaluated at 25,600 evenly space initial state $\delta$ and $\omega$ for ten different $\alpha$ values. The resulting ROA is given in \cref{Res_ROA}.

Based on the results presented in \cref{Res_ROA}, we observed that the Re-PINN accurately classified the PLL system's stable and unstable initial states compared to the ROM proposed in \cite{Kazem_ROM}. Notably, the Re-PINN achieved this in 10 minutes. Even when considering the dataset generation and the training, the Re-PINN only took less than one hour, while the ROM required over two hours to evaluate the system trajectories using the HPC cluster. We attribute this performance improvement to our use of GPUs and CUDA for training the Re-PINN.

Traditionally, CPUs were used for computationally intensive tasks, while GPUs were reserved for graphic rendering tasks. However, unlike CPUs, which have a limited number of processing cores, GPUs have hundreds of less powerful cores with high memory bandwidth. This makes GPUs extremely useful for highly parallelizable tasks like ML training. By utilizing GPUs for Re-PINN training, we achieved a speed-up of 10 to 20 times compared to CPUs in the HPC, making online training and deployment of the Re-PINN competitive with ROM.

\begin{figure*}
  \centerline{\includegraphics[scale=.35]{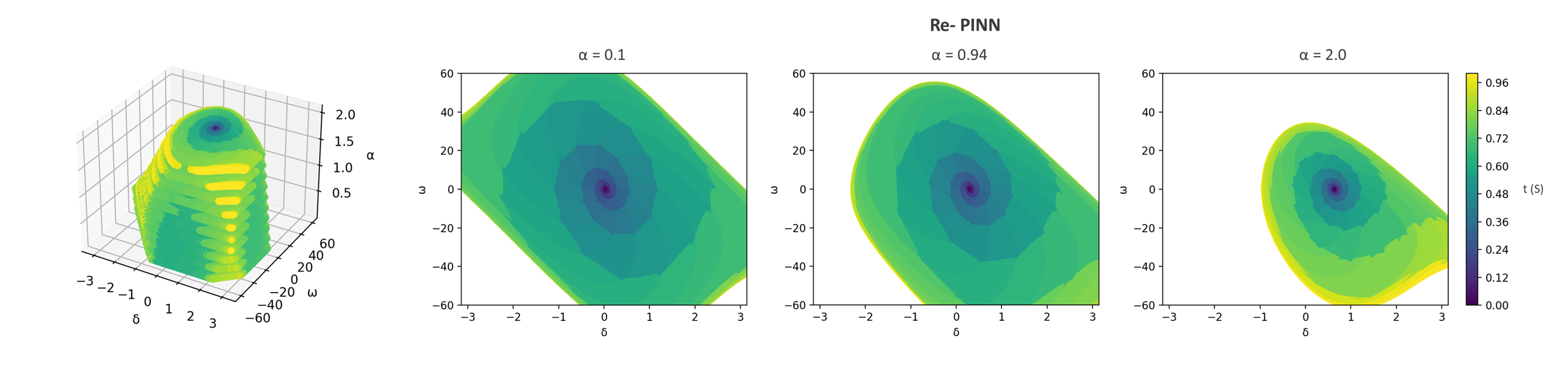}}
  \caption{Region of attraction predicted by the Re-PINN for 5 million evenly spaced random initial states.}
  \label{Res_ROA_5M}
\end{figure*}

Additionally, the Re-PINN enabled the rapid assessment of the ROA for a significantly larger number of points than the ROM. Using Re-PINN, we evaluated the PLL trajectory for 5 million evenly spaced random initial states in under half an hour, a task that would have taken the ROM over two days. The resulting ROA is given in \cref{Res_ROA_5M}. For a more detailed depiction of the ROA, please refer to \cite{Code_git} for the high-definition images. 

\section{Conclusion}

In this paper, we present a novel Recurring Physics-Informed Neural Network (Re-PINN) architecture to (i) accurately predict the nonlinear transient dynamics of a PLL controller under fault with limited labeled training data, and (ii) at much faster time scales than conventional simulation approaches. To evaluate the performance of our proposed Re-PINN algorithm, we compared it against a Reduced Order Model (ROM) for a renewable generator with an SRF-PLL controller with varying grid impedance. The results demonstrate that the Re-PINN can accurately approximate the trajectories for varying grid impedance. Leveraging the GPU acceleration for the Re-PINN training, we demonstrate that the online training and deployment of the Re-PINN algorithm is orders of magnitude faster than an existing ROM approximation. We show that the Re-PINN algorithm can generate a detailed Region of Attraction for the nonlinear dynamic system, with five million different initial conditions, in just half an hour. In contrast, the ROM approximation would have taken more than two days to compute. In the future, this work will focus on the scalability of this approach, expanding to include more system and control parameters in the NN approximator

\bibliographystyle{IEEEtran}
{
\bibliography{main}}
\end{document}